\documentclass[%
 reprint,
 superscriptaddress,
nofootinbib,
 amsmath,amssymb,
 aps,
 prl,
floatfix,
twocolumn,
notitlepage
]{revtex4-2}
\usepackage{bbm}
\usepackage{lipsum}
\usepackage{graphicx}
\usepackage{dcolumn}
\usepackage{subfigure}
\usepackage{bm}
\usepackage{breakurl}
\usepackage{enumitem}
\usepackage{ragged2e}
\usepackage[normalem]{ulem}
\usepackage{multirow}
\usepackage[multidot]{grffile}

\usepackage{hyperref}
\usepackage{geometry}
\usepackage{comment}
\usepackage{algorithm}
\usepackage{algpseudocode}

\newcommand{\be}{\begin{equation}}
\newcommand{\ee}{\end{equation}}

\def\bc{\begin{center}}
\def\ec{\end{center}}
\def\bea{\begin{eqnarray}}
\def\eea{\end{eqnarray}}

\newcommand{\abs}[1]{\left|#1\right|}

\def\multiset#1#2{\ensuremath{\left(\kern-.3em\left(\genfrac{}{}{0pt}{}{#1}{#2}\right)\kern-.3em\right)}}

\usepackage{dcolumn}
\usepackage[table]{xcolor}
\usepackage{booktabs}

\newcommand{\probP}{\text{I\kern-0.15em P}}
\begin{document}

\title{Belief propagation for finite networks using a symmetry-breaking source node}

\author{Seongmin Kim}
\affiliation{Institute of Data Science, University of Hong Kong, Hong Kong SAR, China}

\author{Alec Kirkley}
\email{alec.w.kirkley@gmail.com}
\affiliation{Institute of Data Science, University of Hong Kong, Hong Kong SAR, China}
\affiliation{Department of Urban Planning and Design, University of Hong Kong, Hong Kong SAR, China}
\affiliation{Urban Systems Institute, University of Hong Kong, Hong Kong SAR, China}

\date{\today}

\begin{abstract}
Belief Propagation (BP) is an efficient message-passing algorithm widely used for inference in graphical models and for solving various problems in statistical physics.
However, BP often yields inaccurate estimates of order parameters and their susceptibilities in finite systems, particularly in sparse networks with few loops. 
Here, we show for both percolation and Ising models that fixing the state of a single well-connected ``source'' node to break global symmetry substantially improves inference accuracy and captures finite-size effects across a broad range of networks, especially tree-like ones, at no additional computational cost.
\end{abstract}

\maketitle

Message passing enables efficient inference in probabilistic graphical models of complex systems~\cite{Pearl1988Probabilistic, Wainwright2007Graphical, Koller2010Probabilistic}. This approach has led to advances across diverse fields, including machine learning~\cite{Kschischang2001Factor, Jordan2004Graphical, Alpaydin2010Introduction}, epidemiology~\cite{Karrer2010Message, Altarelli2014Bayesian}, social network analysis~\cite{Hastings2006Community, Decelle2011Inference, Nguyen2013Budgeted, Pei2020Influencer}, and statistical physics~\cite{Nishimori2001Statistical, Zdeborova2016Statistical, Mezard2009Information}. Central to message passing is Belief Propagation (BP), also known as the cavity method, which efficiently computes marginal probabilities by decomposing global inference problems as the exchange of local messages between nodes and their neighbors. 
BP is exact on acyclic (tree) networks and often provides good approximations on cyclic (loopy) networks that are locally tree-like~\cite{Murphy1999Loopy, Mezard2009Information}.

In statistical physics, the application of BP to phase transition models, such as percolation~\cite{Karrer2014Percolation, Allard2019Accuracy, Cantwell2019Message, Cantwell2023Heterogeneous, Mann2023Belief, Newman2023Message} and the Ising model~\cite{Mooij2005Properties, Dorogovtsev2008Critical, Mezard2009Information, Yoon2011Beliefpropagation, Kirkley2021Belief, Newman2023Message}, has yielded highly accurate predictions in many large real-world networks. However, this narrative of success holds a profound exception: BP's ability to describe phase transitions catastrophically breaks down on the very structures where it is theoretically exact---tree networks---creating a paradox~\cite{Allard2019Accuracy}. This failure persists on \textit{almost-trees}, networks that are only slightly more connected.

This paradox arises because the system is symmetric and finite, and thus lacks true spontaneous symmetry breaking (SSB) in theory~\cite{Dorogovtsev2005Correlations, Fraser2016Spontaneous}. Although finite systems may appear to exhibit SSB and nontrivial order parameters, these are practical measurements rather than quantities defined in the thermodynamic limit.
For example, percolation strength refers to the fractional size of the infinite cluster, which exists only in infinitely large systems~\cite{Stauffer1994Introduction}; similarly, the average magnetization in a finite spin system vanishes if measured over an infinite time as thermal fluctuations restore symmetry~\cite{Newman1999Monte}. Thus, theoretical order parameters are zero in finite systems.

BP is indeed exact on trees and correctly predicts the absence of SSB~\cite{Allard2019Accuracy}. However, this theoretical consistency on trees makes BP incompatible with practical metrics used in finite systems, such as the fraction of the largest cluster in percolation or the absolute magnetization in the Ising model. This discrepancy highlights the need for a modified BP approach to accurately infer these practical metrics in finite systems. Ref.~\cite{Allard2019Accuracy} partially addressed this ``tree-like network catastrophe'' by adding a clique to the tree to induce a nontrivial solution, but the resulting accuracy remains poor.

Here, we introduce the Source-Node BP (SNBP) method, a simple modification to the BP framework that explicitly breaks system symmetry by designating a single source node. This enables BP to accurately approximate order parameters and susceptibilities while maintaining computational efficiency. We show that SNBP significantly reduces finite-size errors across various networks, especially tree-like ones, and closely matches ground-truth results from Monte Carlo (MC) simulations. SNBP also outperforms conventional BP and the naive mean-field approximation (MFA) in many real-world networks, offering a practical inference method for finite systems with global symmetries.

We demonstrate our framework on two canonical phase transition models, bond percolation and the Ising model, defined on an undirected graph $\mathcal{G}=(\mathcal{V},\mathcal{E})$ with $N$ nodes and $M$ edges.

\textit{Bond percolation}---Each bond $(i,j)\in\mathcal{E}$ is independently occupied with probability $p$. In the large-$N$ limit, the model exhibits a phase transition at a critical probability $p_c$: for $p < p_c$, only finite clusters exist, while for $p > p_c$, an infinite cluster spans a nonzero fraction of the network.
The theoretical order parameter (percolation strength) is the probability $P_\infty$ that a randomly chosen node belongs to the infinite cluster, which is strictly zero in any finite system. In practice, for a finite graph, we measure the practical order parameter as the fractional size of the largest cluster, $S_1 = {|\mathcal{C}_{\mathrm{max}}|}/{N}$, where $|\mathcal{C}_{\mathrm{max}}|$ is the number of nodes in the largest cluster, $\mathcal{C}_{\mathrm{max}}$~\cite{Stauffer1994Introduction}. In the large-$N$ limit, $S_1\to P_\infty$.
The theoretical susceptibility $\chi_\mathrm{true} =\sum_{|\mathcal{C}|<\infty} |\mathcal{C}|^2 /N$  measures the mean finite cluster size, where $|\mathcal{C}|$ is the size of cluster $\mathcal{C}$. In finite systems, the infinite cluster is absent, and the direct calculation of $\chi_\mathrm{true}$ yields no peak. Therefore, the practical susceptibility is defined by excluding the largest cluster, treating it as the infinite cluster: $\chi_\mathrm{practical} = \sum_{\mathcal{C} \neq \mathcal{C}_{\mathrm{max}}} |\mathcal{C}|^2 /N$~\cite{Lee1996Universal}.

\textit{Ising model}---Each node $i \in \mathcal{V}$ carries a spin $\sigma_i = \pm 1$. In zero field, the Hamiltonian is $\mathcal{H}(\boldsymbol{\sigma}) = -\sum_{(i,j) \in \mathcal{E}} \sigma_i \sigma_j$.
The theoretical order parameter is the average magnetization, $m \equiv \sum_{i=1}^N \sigma_i /N$. In the large-$N$ limit, the system exhibits SSB below a critical temperature $T_c$, and the equilibrium magnetization $\langle m \rangle$ becomes nonzero. However, for any finite system, thermal fluctuations cause global spin flips, restoring the symmetry and ensuring $\langle m \rangle = 0$ at all temperatures. Therefore, the absolute magnetization, $\langle |m| \rangle$, is used as the practical order parameter in finite systems; it is nonvanishing for $T < T_\mathrm{c}$ and converges to $|\langle m \rangle|$ as $N \to \infty$~\cite{Newman1999Monte}.
The magnetic susceptibility for each order parameter is given by: $\chi_\mathrm{true} = \beta N \left( \langle m^2 \rangle - \langle m \rangle^2 \right)$, and $\chi_\mathrm{practical} = \beta N \left( \langle m^2 \rangle - \langle |m| \rangle^2 \right)$, where $\beta \equiv 1/T$~\cite{Landau2014Guide}. In finite systems, $\langle m \rangle = 0$, so $\chi_\mathrm{true}$ simplifies to $\beta N \langle m^2 \rangle$.

We introduce the SNBP framework for these two models. SNBP breaks global symmetry by clamping the state of a designated source node $x$ throughout BP iterations, assuming $x$ is always connected to the infinite cluster in percolation or has a fixed spin direction (e.g., positive) in the Ising model. The marginal probability of interest for each system is then the probability that a node is connected to $x$ (percolation) or has a spin aligned with $\sigma_x$ (Ising). Consequently, the order parameters become the fraction of nodes in $x$'s cluster (percolation) or magnetization aligned with $\sigma_x$ (Ising). The source node $x$ can be selected arbitrarily but different choices provide different accuracy improvements. A simple heuristic that works well in practice is to select the highest-degree node as $x$, since it often belongs to the largest cluster or aligns with the dominant magnetization. This facilitates reliable approximations to practical order parameters in finite systems.

Algorithmically, SNBP is distinguished from conventional BP by the inclusion of Kronecker delta terms that enforce a fixed state on the source node. Removing these terms recovers the conventional BP equations (see Supplemental Material for details).
Throughout, we denote by $\mathcal{N}_i$ the set of neighbors of node $i$, and by $\mathcal{N}_j\setminus i$ the set of neighbors of $j$ excluding $i$. The cluster containing the source node $x$ is denoted by $\mathcal{C}(x)$.

\textit{SNBP for percolation}---We derive the SNBP equations by extending conventional BP for percolation~\cite{Karrer2014Percolation, Allard2019Accuracy, Cantwell2019Message}. Let $\mu^{(x)}_{i \gets j}$ denote the probability that node $j$ belongs to $\mathcal{C}(x)$ when $i$ is removed, and $\mu^{(x)}_i$ the probability that node $i$ belongs to $\mathcal{C}(x)$. Since the source node is always connected to $\mathcal{C}(x)$, $\mu^{(x)}_{i\gets x}=1$ for $i\in\mathcal{N}_x$ and $\mu^{(x)}_{x}=1$. This modification can be incorporated into the message update rule:
\begin{align}
\mu^{(x)}_{i\gets j} &= 1-(1-\delta_{jx})\prod_{k\in\mathcal{N}_j\setminus i} \left(1-p\,\mu^{(x)}_{j\gets k}\right). \label{eq:1_mu_ij}
\end{align}
The expected fraction of $\mathcal{C}(x)$ is then $\langle{S(x)}\rangle_{\mathrm{SNBP}}\equiv\langle{|\mathcal{C}(x)|}\rangle_{\mathrm{SNBP}} /N = \sum_{i=1}^N \mu^{(x)}_i /N$ where the marginal probability $\mu^{(x)}_i$ satisfies 
\begin{align}
\mu^{(x)}_i  = 1 - (1-\delta_{ix})\prod_{j\in\mathcal{N}_i} \left(1-p\,\mu^{(x)}_{i\gets j}\right). \label{eq:2_mu_i}
\end{align}

The message-passing equation for determining the susceptibility in the SNBP framework can be derived from linear response analysis~\cite{Welling2004Linear, Mezard2009Constraint, Karrer2014Percolation} and is given by
\begin{align}
\chi^{(x)}_{i\gets j} &= \left[1 + \sum_{k\in\mathcal{N}_j\setminus i} \frac{p\,\chi^{(x)}_{j\gets k}}{1- p\,\mu^{(x)}_{j\gets k}}\right] \left(1-\mu^{(x)}_{i\gets j}\right). \label{eq:3_chi_ij}
\end{align}
The expected global susceptibility, representing the mean size of non-$\mathcal{C}(x)$ clusters, is then $\chi^{(x)}_\mathrm{SNBP} = \sum_{i=1}^N \chi^{(x)}_i /N$ where 
\begin{align}
\chi^{(x)}_i = \left[1 + \sum_{j\in\mathcal{N}_i} \frac{p\,\chi^{(x)}_{i\gets j}}{1-p\,\mu^{(x)}_{i\gets j}}\right] \left(1-\mu^{(x)}_i\right).    \label{eq:4_chi_i}
\end{align}

\textit{SNBP for the Ising model}---Extending the conventional BP framework~\cite{Mezard2009Information, Nguyen2017Inverse, Newman2023Message}, we derive the SNBP equations for the Ising model. Let $h^{(x)}_{i \gets j}$ denote the effective field from spin $j$ to spin $i$ when spin $x$ is fixed to $\sigma_x=+1$. The expected spin of $x$ is fixed to $m_x=1$ by setting $h^{(x)}_{x\gets i}=\infty$ for $i\in\mathcal{N}_x$.
This constraint can be incorporated into the message update rule:
\begin{align}
\tanh(\beta h^{(x)}_{i\gets j}) = & \;\delta_{ix} + (1-\delta_{ix}) \, \tanh(\beta) \nonumber \\
& \times \tanh\left(\sum_{k\in\mathcal{N}_j \setminus i} \beta h^{(x)}_{j\gets k}\right). \label{eq:5_h_ij}
\end{align}
The expected global magnetization aligned with $\sigma_x$ is then $\langle{m}\rangle^{(x)}_{\mathrm{SNBP}} = \sum_{i=1}^N{m^{(x)}_i} /N$ where
\begin{align}
m^{(x)}_i = \delta_{ix} + (1-\delta_{ix})\tanh\left(\sum_{j\in\mathcal{N}_i} \beta h^{(x)}_{i\gets j}\right)    \label{eq:6_m_i}
\end{align}
is the probability that $\sigma_i$ aligns with $\sigma_x$.

Following the approach in Refs.~\cite{Welling2004Linear, Mezard2009Constraint, Aurell2010Dynamics}, we derive the susceptibility propagation under a uniform external field:
\begin{align}
q^{(x)}_{i\gets j} = Q^{(x)}_{ij} \Biggl( 1 + \sum_{k \in \mathcal{N}_j \setminus i} q^{(x)}_{j\gets k} \Biggr), \label{eq:7_q_ij}
\end{align}
where
\begin{align}
Q^{(x)}_{ij} = \frac{\tanh(\beta)}{\mathrm{sech}^2(\beta h^{(x)}_{i\gets j})}  \left[ 1 - \frac{\tanh^2 (\beta h^{(x)}_{i\gets j})}{\tanh^2(\beta)} \right].  \label{eq:8_Q_ij}  
\end{align}
The global susceptibility is then $\chi^{(x)}_\mathrm{SNBP} = \sum_{i=1}^N \chi^{(x)}_i /N$, where 
\begin{align}
\chi^{(x)}_i = \beta \left[1 - (m^{(x)}_i)^2\right] \Biggl( 1 + \sum_{j \in \mathcal{N}_i} q^{(x)}_{i\gets j} \Biggr). \label{eq:9_chi_i}
\end{align}
Detailed derivations of Eqs.~\eqref{eq:1_mu_ij}--\eqref{eq:9_chi_i} are provided in the Supplemental Material. For direct comparison with percolation, Ising model results are plotted against the equivalent bond occupation probability $p = 1 - e^{-2\beta}$ from the random cluster representation~\cite{Wu1982Potts}.

We also apply this source-node concept to extend MFA and MC simulations. We call these extensions SNMFA and SNMC respectively. In SNMFA, the state of $x$ is fixed during mean-field updates, and in SNMC, it is fixed during Monte Carlo sampling (see Supplemental Material for details). All variants retain computational efficiency similar to their conventional forms.

\begin{figure}
\includegraphics[width=\columnwidth,
                    trim={0.18cm 0.12cm 0cm 0cm}, 
                    clip,
                    ]{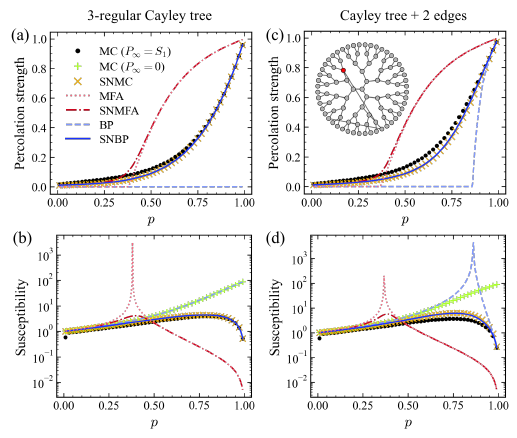} 
\caption{\label{fig:1} \RaggedRight 
(a),(c) Percolation strength $P_{\infty}$ and (b),(d) susceptibility $\chi$ as functions of occupation probability $p$ for (a),(b) a 3-regular Cayley tree ($N=94$, $M=93$) and (c),(d) the same tree with two additional edges.
Symbols indicate results from MC and SNMC.
Curves show message-passing results inferred by MFA, SNMFA, BP, and SNBP. 
The curve `MC ($P_{\infty}=S_1$)' uses $S_1$ as the order parameter and $\chi_\mathrm{practical}$ as the susceptibility, while
`MC ($P_{\infty}=0$)' computes $\chi_\mathrm{true}$.
The red node in each network marks the source node, chosen to be of the highest degree.
}
\end{figure}

\textit{Trees and almost-trees}---We first evaluate the performance of these source-node inference methods on synthetic networks under the bond percolation model, starting with a perfect tree and its perturbations with a few added edges. For the perfect tree, we use a 3-regular Cayley tree of depth 5 ($N=94$, $M=93$), where internal nodes have degree 3 and leaves have degree 1.

On the perfect Cayley tree, conventional BP stably converges to a trivial solution, correctly capturing the absence of SSB: $\langle{P_\infty}\rangle_{\mathrm{BP}} = 0$ (Fig.~1(a)). The susceptibility computed by conventional BP, $\chi_{\mathrm{BP}}$, increases monotonically with $p$ and exactly matches ground-truth MC results, $\chi_\mathrm{true}$, obtained by assuming $P_\infty=0$ (Fig.~1(b)). However, these results are useless for predicting the practical phase-transition metrics, which estimate $P_\infty$ using $S_1=\abs{\mathcal{C}_{\mathrm{max}}}/N$.

Adding a single edge between a randomly chosen node pair in the tree causes conventional BP to exhibit a spurious critical point at $p_\mathrm{c}^{\mathrm{BP}}=1$ (Supplemental Material Figs.~S1(a),(b)). This occurs because BP treats the resulting cycle as an infinite 1D chain~\cite{Allard2019Accuracy} with a known critical point at $p=1$ for percolation ($T=0$ for the Ising model). As a result, $\langle{P_\infty}\rangle_{\mathrm{BP}}$ converges to zero for $p<1$, as it does on a tree, but becomes under-constrained at $p_\mathrm{c}^{\mathrm{BP}}=1$.
$\chi_{\mathrm{BP}}$ deviates from $\chi_\mathrm{true}$ in MC simulations and diverges to infinity at $p_\mathrm{c}^{\mathrm{BP}}=1$.

When two or more edges are added to the tree, $p_{\mathrm{c}}^{\mathrm{BP}}$ shifts below unity. For $p > p_{\mathrm{c}}^{\mathrm{BP}}$, BP admits both the trivial and a spurious nontrivial solution; however, as the trivial solution is unstable, BP reliably converges to the spurious solution under asymmetric initialization. Consequently, $\langle P_\infty \rangle_{\mathrm{BP}}$ becomes nonzero for $p > p_{\mathrm{c}}^{\mathrm{BP}}$ (Fig.~1(c), Fig.~S1(c)), and $\chi_{\mathrm{BP}}$ displays a divergent peak at $p_{\mathrm{c}}^{\mathrm{BP}}$ (Fig.~1(d), Fig.~S1(d)).
Although adding a few cycles to the tree improves BP's estimation of practical metrics, it is still inadequate near and below $p_{\mathrm{c}}^{\mathrm{BP}}$.

In contrast, SNBP dramatically reduces this gap between theory and numerical experiments by explicitly breaking the symmetry. It admits only a nontrivial solution, which accurately captures the nonzero $\langle{S_1}\rangle_{\mathrm{MC}}$ and the smooth peak of $\chi_{\mathrm{MC}}$ (Fig.~1). By construction, SNBP exactly matches SNMC whenever the removal of the source node yields a perfect tree (Fig.~1(a),(b), Fig.~S1(a),(b)).
This exactness is rooted in the same principle as Conditioned BP (CBP), which achieves exactness by clamping nodes to remove all cycles~\cite{Pearl1986Fusion, Bidyuk2007Cutset, Eaton2009Choosing}.
However, unlike SNBP, CBP does not address the issue of global symmetry and its computational complexity grows exponentially with system size due to scanning over all possible clamped values.
Even when an extra cycle remains, SNBP still closely predicts SNMC (Fig.~1(c),(d)).
Since the highest-degree node often lies in the largest cluster, SNMC---and thus SNBP---reliably approximates the practical metrics of conventional MC.

Although the naive MFA predicts a nonzero order parameter and a susceptibility peak even on a perfect tree (Fig. 1(a),(b)), these predictions deviate significantly from MC results. This inaccuracy persists despite the addition of several cycles (Fig.~1(c),(d), Fig.~S1(c),(d)). Its source-node variant, SNMFA, is similarly inaccurate, except for slightly reduced finite-size errors at small $p$ and non-diverging susceptibility. Overall, SNBP clearly outperforms both MFA and SNMFA in accuracy for practical metrics on trees and almost-trees.

\begin{figure}
\includegraphics[width=\columnwidth,
                    trim={0.18cm 0.12cm 0cm 0cm}, 
                    clip,
                    ]{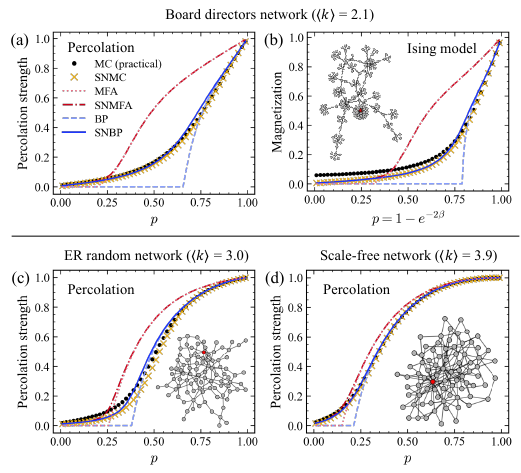}
\caption{\label{fig:2}
Inference on locally tree-like networks. Order parameters for the percolation model are shown in (a), (c), and (d), and for the Ising model in (b).
Simulations were run for the Norwegian board directors network ($N=179$, $M=184$), an
Erdos-Renyi random graph ($N=94$, $M=139$),
and a scale-free network ($N=80$, $M=156$) generated using the Barabasi-Albert model with $k=2$ out-edges per arriving node.
The red node in each network marks the source node.
}
\end{figure}

\textit{Locally tree-like networks}---The high accuracy of SNBP, demonstrated on a Cayley tree and its perturbations, generalizes to various locally tree-like networks. On the real-world Norwegian board directors network, an almost-tree, SNBP shows superior accuracy for the order parameter (Figs.~2(a),(b)) and susceptibility (Figs.~S2(a),(b)). A minor deviation occurs only for the Ising model at high temperature (small $p=1-e^{-2\beta}$), where $\langle m \rangle_{\mathrm{SNBP}} \to 1/N$ but $\langle|m|\rangle_{\mathrm{MC}} \to (2/(\pi N))^{1/2}$ (derived from the one-dimensional random-walk result of~\cite{Feller1950Introduction}). In contrast, SNBP is exact for percolation in this limit: as $p\to0$, both $\langle \mathcal{C}(x)\rangle_{\mathrm{SNBP}}$ and $\langle S_1\rangle_{\mathrm{MC}}$ approach $1/N$.

Even when networks contain many cycles---as quantified by a large cyclomatic number $c(\mathcal{G}) \equiv M-N+1$---SNBP maintains high accuracy if the network is locally tree-like. This accuracy stems from the established principle that conventional BP provides highly accurate inference in such networks~\cite{Mezard2009Information, Newman2023Message}.
Our simulations show that SNBP outperforms the other inference methods in ER random networks (Fig.~2(c), Fig.~S2(c)) and scale-free networks (Fig.~2(d), Fig.~S2(d)).
Although their cyclomatic numbers are not small, the sparsity of short cycles enables SNBP to remain accurate.

\begin{figure}
\includegraphics[width=\columnwidth,
                    trim={0.18cm 0.12cm 0cm 0cm}, 
                    clip,
                    ]{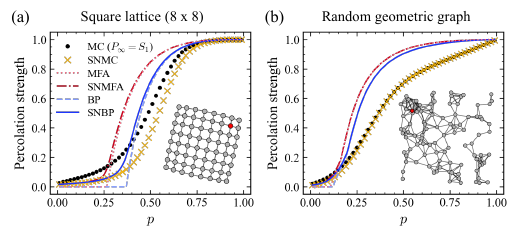}
\caption{\label{fig:3} \RaggedRight 
Inference for the percolation strength on spatial networks:
(a) $8\times8$ square lattice, 
(b) random geometric graph ($N=100$, $\langle{k}\rangle = 5.9$).
Both BP and SNBP fail to reproduce MC results, showing significant deviations over a wide range of $p$.
Red node in networks marks source node.
}
\end{figure}

\textit{Spatial networks}---In networks that are not locally tree-like and contain many short cycles, such as spatial networks including lattices (Fig.~3(a)) and random geometric graphs (Fig.~3(b)), both BP and SNBP deviate significantly from MC results. Although SNBP is more accurate than BP for $p < p_{\mathrm{c}}^{\mathrm{BP}}$, its overall accuracy is still limited by errors from cycles---a well-known issue in loopy BP. These cycle-driven errors can be overcome by more computationally intensive algorithms that explicitly account for neighborhood correlations~\cite{Kirkley2021Belief, Cantwell2019Message, Cantwell2023Heterogeneous, Mann2023Belief}.

\begin{figure}
\includegraphics[width=\columnwidth,
                    trim={0.18cm 0.12cm 0cm 0cm}, 
                    clip,
                    ]{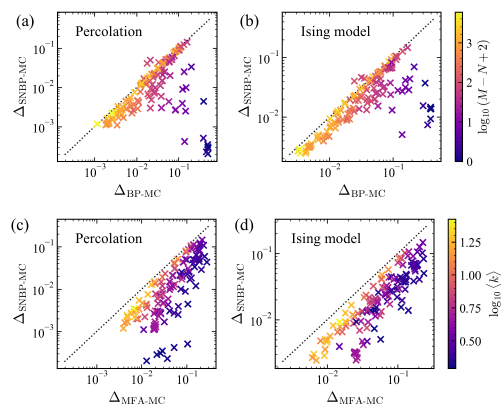}
\caption{\label{fig:4} \RaggedRight 
Comparison of SNBP errors with BP and MFA errors for (a),(c) percolation model and (b),(d) the Ising model on 139 real networks.
(a),(b) SNBP outperforms BP: $\Delta_{\mathrm{SNBP-MC}}$ is generally smaller than $\Delta_{\mathrm{BP-MC}}$, particularly for small cyclomatic numbers.
(c),(d) SNBP outperforms MFA: $\Delta_{\mathrm{SNBP-MC}}$ is almost always smaller than  $\Delta_{\mathrm{MFA-MC}}$. The gap increases as $\langle{k}\rangle$ decreases.
}
\end{figure}

\textit{Real-world networks}---To systematically evaluate the performance of SNBP, we benchmark its accuracy against conventional BP and MFA on 139 real-world networks (listed in Supplemental Materials). For each network, we extract the largest connected component and convert it to a simple graph by removing directionality, duplicate edges, and self-loops. Accuracy is quantified by the mean absolute error in the order parameter relative to conventional MC simulations, $\Delta_{\mathrm{Method-MC}}$, which is calculated as the area between the curves from $p=0$ to $1$ (see Supplemental Materials).

For both the percolation (Fig.~4(a)) and Ising (Fig.~4(b)) models, SNBP consistently outperforms conventional BP across the real-world networks. The improvement is most pronounced in networks with a low cyclomatic number, for which conventional BP suffers from the aforementioned ``tree-like network catastrophe'' and can provide even less accurate results than MFA (Figs.~S4(a),(b)).

Unlike conventional BP, SNBP outperforms MFA across all tested networks (Figs. 4(c),(d)). The performance gap increases with decreasing $\langle k \rangle$, highlighting SNBP's particular advantage in weakly connected networks where mean-field assumptions break down.

\textit{Discussion}---Here, we address a fundamental yet often overlooked challenge in applying BP to finite, symmetric phase transition models: BP's exactness on trees, which correctly predicts the absence of spontaneous symmetry breaking, prevents it from effectively capturing established practical order parameters~\cite{Allard2019Accuracy}.
To bridge this gap between theory and practice, we propose SNBP, which breaks global symmetry by fixing a single source node. Our tests on percolation and Ising models show that SNBP accurately infers practical order parameters and their susceptibilities in a broad range of real and synthetic networks. SNBP most significantly outperforms conventional BP in networks with few cycles and naive MFA in sparse networks. Importantly, this enhanced accuracy is achieved without sacrificing the computational efficiency of standard BP.

Opportunities remain for further improvement and broader application. Like conventional BP, SNBP's accuracy is limited by the presence of short cycles. This can be mitigated by integrating the source-node method with more sophisticated BP algorithms for loopy graphs~\cite{Kirkley2021Belief, Cantwell2019Message, Cantwell2023Heterogeneous, Mann2023Belief}, albeit at the cost of increased computational expense.
Additionally, in networks lacking a clear hub, selecting the highest-degree node as the source may not be optimal. Future research could explore alternative strategies for choosing one or more source nodes to improve accuracy.

Moreover, SNBP shows promise for extension to other symmetric phase transition models, including the $q$-state Potts model~\cite{Wu1982Potts}, XY model~\cite{Lupo2017Approximating}, and Heisenberg model~\cite{DelBono2024Analytical}. Beyond these, the core principle of SNBP---strategically fixing node or edge values for practical message-passing approximations---could potentially be adapted to diverse inference tasks on networks, such as community detection~\cite{Hastings2006Community, Decelle2011Inference}, as well as various deep learning tasks~\cite{Baldassi2020Shaping, Lucibello2022Deep, Huang2021Statistical}.


\bibliographystyle{numeric}
\bibliography{arXiv/_bib_KimKirkley2025}


%
%

\clearpage

\onecolumngrid


\begin{center}
  \textbf{\large Supplemental Material for: \\ \vspace{0.25cm}
Symmetry-Breaking Source-Node Belief Propagation for Finite Networks \vspace{0.25cm}} \\[.2cm]

Seongmin Kim,$^{1}$, Alec Kirkley,$^{1,2,3}$ \\ [.1cm]
  {\itshape ${}^1$Institute of Data Science, University of Hong Kong, Hong Kong SAR, China \\
            ${}^2$Department of Urban Planning and Design, University of Hong Kong, Hong Kong SAR, China \\
            ${}^3$Urban Systems Institute, University of Hong Kong, Hong Kong SAR, China \\
            }
\end{center}

\setcounter{equation}{0}
\setcounter{figure}{0}
\setcounter{table}{0}
\setcounter{page}{1}
\setcounter{section}{0}
\renewcommand{\theequation}{S\arabic{equation}}
\renewcommand{\thefigure}{S\arabic{figure}}
\renewcommand{\thetable}{S\arabic{table}}
\renewcommand{\thepage}{S\arabic{page}}
\renewcommand{\thesection}{S\arabic{section}}
\renewcommand{\bibnumfmt}[1]{[#1]}
\renewcommand{\citenumfont}[1]{#1}

\section{Supplemental Figures}

\begin{figure}[h]
\includegraphics[scale = 1.05,
                    trim={0.18cm 0.12cm 0cm 0cm}, 
                    clip,
                    ]{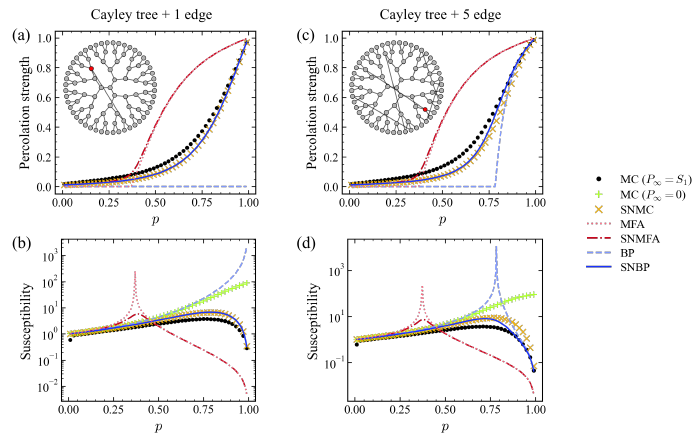} 
\caption{\label{supfig:1} \RaggedRight 
Supplemental figure for Fig.~1. (a),(c) Percolation strength $P_{\infty}$ and (b),(d) susceptibility $\chi$ as functions of occupation probability $p$ for 
(a),(b) a 3-regular Cayley tree ($N=94$) with one additional edge and
(c),(d) the same tree with five additional edges.
`MC ($P_{\infty}=S_1$)' uses $S_1$ as the order parameter and $\chi_\mathrm{practical}$ as the susceptibility.
`MC ($P_{\infty}=0$)' computes $\chi_\mathrm{true}$.
Red node in networks marks source node.
}
\end{figure}

\begin{figure}[h]
\includegraphics[scale = 1.05,
                    trim={0.18cm 0.12cm 0cm 0cm}, 
                    clip,
                    ]{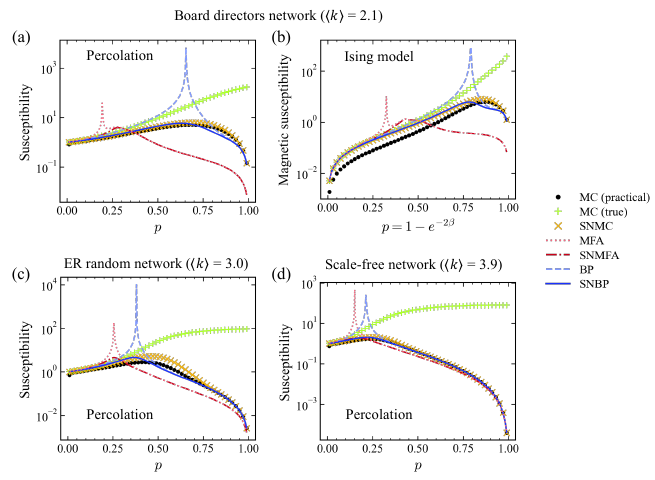}
\caption{\label{supfig:2}
Supplemental figure for Fig.~2. Inference on locally tree-like networks. Susceptibilities for the percolation model in (a), (c), and (d), and for the Ising model in (b).
Networks: Norwegian board directors network ($N=179$, $M=184$; a, b),
ER random network ($N=94$, $M=139$; c),
and scale-free network ($N=80$, $M=156$; d).
`MC (practical)' uses $S_1$ (percolation) and $\langle|M|\rangle$ (Ising) as the order parameter and $\chi_\mathrm{practical}$ as the susceptibility.
`MC (true)' computes $\chi_\mathrm{true}$.
}
\end{figure}

\begin{figure}[h]
\includegraphics[scale = 1.05,
                    trim={0.18cm 0.12cm 0cm 0cm}, 
                    clip,
                    ]{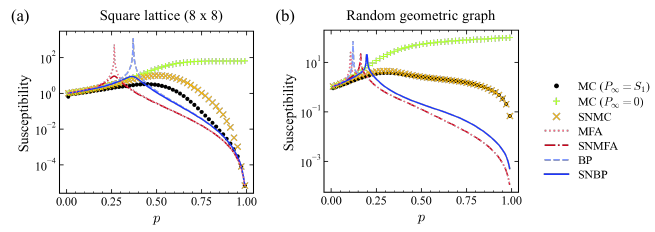}
\caption{\label{supfig:3} \RaggedRight 
Supplemental figure for Fig.~3. Inference for percolation susceptibility on spatial networks:
(a) $8\times8$ square lattice, 
(b) random geometric graph ($N=100$, $\langle{k}\rangle = 5.9$).
Both BP and SNBP fail to reproduce MC results, showing significant deviations over a wide range of $p$.
}
\end{figure}

\begin{figure}[h]
\includegraphics[scale = 1.1,
                    trim={0.18cm 0.12cm 0cm 0cm}, 
                    clip,
                    ]{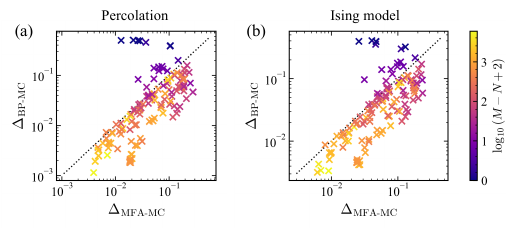}
\caption{\label{supfig:4} \RaggedRight 
Supplemental figure for Fig.~4. Comparison of BP errors with MFA errors for (a) the percolation model and (b) the Ising model on 139 real networks. Conventional BP yields larger errors than MFA on trees or almost-trees: $\Delta_{\mathrm{BP-MC}}>\Delta_{\mathrm{MFA-MC}}$ for small cyclomatic numbers (dark symbols).
}
\end{figure}

\section{Derivation of BP for bond percolation}

Following Refs.~\cite{Karrer2014Percolation, Allard2019Accuracy, Cantwell2019Message}, we derive the conventional BP equations for bond percolation using the cavity method and generating functions, introducing slightly different function definitions. 
Consider a network with nodes $i=1,\dots,N$, and let $\mathcal{N}_i$ denote the set of neighbors of node $i$. Each edge is occupied with probability $p$.

We define $\phi_{i\gets j}(s)$ as the probability that node $j$ belongs to a finite cluster of size $s$ in the absence of $i$.
On a tree graph, $\phi_{i\gets j}(s)$ can be written recursively as
\begin{equation}
\phi_{i\gets j}(s) = \sum_{\{s_k | k\in\mathcal{N}_j\setminus i\}}  \left[ \prod_{k\in\mathcal{N}_j\setminus i} \bigl( (1-p)\,\delta_{0,\,s_k} + p\,\phi_{j\gets k}(s_k) \bigr) \right] \delta_{ s-1, \,\sum_{k\in\mathcal{N}_j\setminus i} s_k }
\label{eq:phi_cavity}
\end{equation}
for $s\geq0$, where $s_k \in \mathbb{Z}_{\geq 0}$ for all $k$.
$\mathcal{N}_j\setminus i$ denotes the neighbors of $j$ excluding $i$, and $\delta$ is the Kronecker delta.
The factor $(1-p)\,\delta_{0,\,s_k}$ accounts for the case where the bond $(j,k)$ is absent ($s_k=0$), while $p\,\phi_{j\gets k}(s_k)$ accounts for the case where the bond is present ($s_k\geq 1$).

Let $\phi_i(s)$ be the probability that node $i$ belongs to a cluster of size $s$, which can be obtained by combining contributions from all neighbors $j\in\mathcal{N}_i$:
\begin{equation}
\phi_i(s) = \sum_{\{s_j | j\in\mathcal{N}_i\}} \left[\prod_{j\in\mathcal{N}_i} \bigl((1-p)\,\delta_{0,\,s_j} + p\,\phi_{i\gets j}(s_j) \bigr)\right] \delta_{s-1,\,\sum_{j\in\mathcal{N}_i}s_j}
\label{eq:phi_node}
\end{equation}
for $s\geq0$. Note that $\phi_{i\gets j}(0) = 0$ and $\phi_{i}(0) = 0$ for all edges and nodes.

We define generating functions $H_{i\gets j}(z) \equiv \sum_{s=1}^\infty \phi_{i\gets j}(s) z^s$ and $H_i(z) \equiv \sum_{s=1}^\infty \phi_i(s) z^s$.
Multiplying Eqs.~\eqref{eq:phi_cavity} and \eqref{eq:phi_node} by $z^s$ and summing over $s$ yields the recursive relations for the generating functions:
\begin{align}
H_{i\gets j}(z)
&= z \prod_{k\in\mathcal{N}_j\setminus i} \bigl( 1-p + p\,H_{j\gets k}(z) \bigr),
\label{eq:mu_ij_recursion}
\end{align}
\begin{align}
H_i(z)
&= z \prod_{j\in\mathcal{N}_i} \bigl( 1-p + p\,H_{i\gets j}(z) \bigr) .
\label{eq:mu_i_recursion}
\end{align}

The conventional BP equations for bond percolation are obtained by evaluating these generating functions at $z = 1$:
\begin{equation}
\mu_{i\gets j} = 1-\prod_{k\in\mathcal{N}_j\setminus i} \bigl(1 - p\mu_{j\gets k}\bigr),
\label{eq:mu_ij_z1}
\end{equation}
\begin{equation}
\mu_i = 1-\prod_{j\in\mathcal{N}_i} \bigl(1 - p\mu_{i\gets j}\bigr),
\label{eq:mu_i_z1}
\end{equation}
where $\mu_{i\gets j} \equiv 1-H_{i\gets j}(1)$ is the probability that node $j$ belongs to the infinite cluster when node $i$ is removed, and $\mu_i \equiv 1-H_i(1)$ is the probability that node $i$ belongs to the infinite cluster. 
The messages sent from leaf nodes $l$ are all 0: $\mu_{i\gets l}=0$ for $i\in\mathcal{N}_l$.
The expected fraction of nodes in the infinite cluster is then given by $\langle{P_\infty}\rangle_{BP} = \sum_{i=1}^N \mu_i /N$.

The source-node belief propagation (SNBP) replaces the concept of the infinite cluster with a specific source node $x$ in the definitions of $\mu_{i\gets j}$ and $\mu_{i}$. Here, the source node is always considered connected to itself. Therefore, the messages sent from the source node $x$ to its neighbor $i\in\mathcal{N}_x$ are
\begin{equation}
     \mu_{i\gets x}=1.
\end{equation}
By the same logic, the marginal probabilities are
\begin{equation}
     \mu_{x}=1.
\end{equation}
This holds even if the source node is a leaf node. We incorporate these boundary conditions in Eq.~\eqref{eq:1_mu_ij} using a Kronecker delta term.

Susceptibility propagation (SusP) can be derived by combining BP with a linear response approach, as outlined in Refs.~\cite{Welling2004Linear, Mezard2009Constraint}.
By differentiating Eqs.~\eqref{eq:mu_ij_recursion} and \eqref{eq:mu_i_recursion} with respect to $z$, we obtain the following expressions: 
\begin{align}
H_{i\gets j}'(z) &= \left[ \frac{1}{z} + \sum_{k\in\mathcal{N}_j\setminus i} \frac{p\,H_{j\gets k}'(z)}{1- p\,\mu_{j\gets k}(z)} \right] H_{i\gets j}(z) ,
\label{eq:mu_ij_prime}
\\
H_i'(z) &= \left[ \frac{1}{z} + \sum_{j\in\mathcal{N}_i} \frac{p\,H_{i\gets j}'(z)}{1- p\,\mu_{i\gets j}(z)} \right] H_i(z) .
\label{eq:mu_i_prime}
\end{align}
Evaluating these equations at $z = 1$ and defining the susceptibility terms as $\chi_i \equiv H_i'(1) = \sum_{s=1}^\infty s \phi_i(s)$ and $\chi_{i\gets j} \equiv H_{i\gets j}'(1) = \sum_{s=1}^\infty s \phi_{i\gets j}(s)$, we derive the SusP equations:
\begin{align}
\chi_{i\gets j} &= \left[ 1 + \sum_{k\in\mathcal{N}_j\setminus i} \frac{p\,\chi_{j\gets k}}{1- p\,\mu_{j\gets k}} \right] (1-\mu_{i\gets j}).
\label{eq:BP_perc_chi_ij}
\\
\chi_i &= \left[ 1 + \sum_{j\in\mathcal{N}_i} \frac{p\,\chi_{i\gets j}}{1-p + p\,\mu_{i\gets j}} \right] (1-\mu_i) ,
\label{eq:BP_perc_chi_i}
\end{align}
These equations are consistent with the formulation in Ref.~\cite{Karrer2014Percolation}, though expressed with different notation. They are also applicable to SNBP when incorporating the boundary conditions $\mu_{i\gets x} = 1$ and $\mu_x = 1$. The global susceptibility is computed as $\chi = \sum_{i=1}^N \chi_i /N$.

\section{Derivation of BP for the Ising model}

Following the framework of Refs.~\cite{Mezard2009Information, Nguyen2017Inverse, Newman2023Message}, we derive the BP equations for the Ising model. In the absence of an external field, the partition function is 
\begin{equation}
Z = \sum_{\{\boldsymbol{\sigma}\}} \prod_{(i,j)} e^{\beta \sigma_i \sigma_j},
\end{equation}
where the spin $\sigma_i \in \{+1, -1\}$ and $(i,j)$ denotes an occupied edge in the graph.
The restricted partition function $Z_{i\gets j}(\sigma_j)$ is defined as the partition function for the subgraph containing node $j$ when node $i$ is removed and spin $j$ is fixed to $\sigma_j$. On a tree, these satisfy the recurrence relation: 
\begin{equation}\label{eq:Zij_relation}
    Z_{i\gets j}(\sigma_j) = \prod_{k\in\mathcal{N}_j \setminus i} \left[ \sum_{\sigma_k=\pm1} e^{\beta \sigma_k \sigma_j} {Z_{j\gets k}(\sigma_k)} \right].
\end{equation}
The partition function for the entire tree with spin $i$ fixed to $\sigma_i$ is given by 
\begin{equation}\label{eq:Zi_relation}
    Z_i(\sigma_i) = \prod_{j\in\mathcal{N}_i} \left[ \sum_{\sigma_j=\pm1} e^{\beta \sigma_i \sigma_j} {Z_{i\gets j}(\sigma_j)} \right].
\end{equation}
The marginal probability for node $i$ to have spin $\sigma_i$ is then 
\begin{equation}
\mu_i(\sigma_i) = \frac{Z_i(\sigma_i)}{\sum_{\sigma = \pm 1} Z_i(\sigma)}.
\end{equation}

To simplify Eq.~\eqref{eq:Zij_relation}, we introduce the cavity field $h_{i\gets j}$, which corresponds to the effective field that spin $i$ experiences from spin $j$:
\begin{equation}\label{eq:hij_def}
    e^{\beta \sigma_i h_{i\gets j}} \propto \sum_{\sigma_j=\pm1} e^{\beta \sigma_j \sigma_i} {Z_{i\gets j}(\sigma_j)},
\end{equation}
where the proportionality constant is independent of $\sigma_i$.
The conventional BP equations for the Ising model are then
\begin{equation}\label{eq:h_relation}
    \tanh{\left(\beta h_{i\gets j}\right)}= \tanh{(\beta)} \cdot \tanh{\left(\sum_{k\in\mathcal{N}_j \setminus i} \beta h_{j\gets k}\right)}.
\end{equation}
For leaf nodes $l$, the messages are $h_{i\gets l} = 0$ for $i \in \mathcal{N}_l$. 
Once the cavity fields $h_{i\gets j}$ are determined via Eq.~\eqref{eq:h_relation}, the expected spin at node $i$, defined as $m_i \equiv \langle \sigma_i \rangle_{\mathrm{BP}} = \mu_i(+1) - \mu_i(-1)$, is given by
\begin{equation}\label{eq:mi_solution}
    m_i = \tanh{\left(\sum_{j\in\mathcal{N}_i} \beta h_{i\gets j}\right)}.
\end{equation}
The expected global magnetization is then $\langle{m}\rangle_{\mathrm{BP}} = \sum_{i=1}^N{m_i} /N$.

In the SNBP, the spin of the source node $x$ is fixed to $m_x=1$ by setting the effective field on $x$ from its neighbors $i\in\mathcal{N}_x$ to 
\begin{equation}
    h_{x\gets i}=\infty.
\end{equation} 
Through Eq.~\eqref{eq:h_relation}, this induces $h_{i\gets x}=1$. This holds even if the source node is a leaf node. We incorporate this boundary condition in Eq.~\eqref{eq:5_h_ij} using Kronecker delta terms.

Following the approach of Refs.~\cite{Mezard2009Constraint, Aurell2010Dynamics}, we derive the SusP equations for the Ising model subject to a uniform external field $h$. The BP equations in this case are:
\begin{align}
    \tanh{\left(\beta h_{i\gets j}\right)} &= \tanh{(\beta)} \cdot \tanh{\left( \beta h + \sum_{k\in\mathcal{N}_j \setminus i} \beta h_{j\gets k} \right)}, \label{eq:h_relation_with_h_ex}
    \\
    m_i &= \tanh{\left( \beta h + \sum_{j\in\mathcal{N}_i} \beta h_{i\gets j} \right)}.\label{eq:mi_solution_with_h_ex}
\end{align}
To obtain the local susceptibility $\chi_i \equiv {d m_i}/{dh}$, we differentiate Eqs.~\eqref{eq:h_relation_with_h_ex} and \eqref{eq:mi_solution_with_h_ex} with respect to $h$. This yields the following recursive equation for $q_{i\gets j} \equiv {d h_{i\gets j}}/{dh}$:
\begin{equation}\label{eq:susp_dh_relation}
    q_{i\gets j} = \frac{\tanh(\beta)}{\mathrm{sech}^2(\beta h_{i\gets j})} \left[ 1 - \frac{\tanh^2(\beta h_{i\gets j})}{\tanh^2(\beta)} \right] \Biggl( 1 + \sum_{k \in \mathcal{N}_j \setminus i} q_{j\gets k} \Biggr).
\end{equation}
By solving Eq.~\eqref{eq:susp_dh_relation} self-consistently for all directed edges, we obtain $q_{i\gets j}$. The local susceptibility is then
\begin{equation}\label{eq:susp_local_chi}
    \chi_i = \beta\,\left(1 - m_i^2\right)\Biggl( 1 + \sum_{j \in \mathcal{N}_i} q_{i\gets j} \Biggr),
\end{equation}
and the global susceptibility is $\chi = \sum_{i=1}^N \chi_i /N$. These equations are also applicable to SNBP when incorporating the boundary conditions $h_{x\gets i} = \infty$ and $m_x = 1$.

\section{Mean-field approximation for bond percolation}

In a homogeneous network where each node $j$ has a large number of neighbors, the influence of any single neighbor $i \in \mathcal{N}_j$ is negligible. Consequently, for bond percolation, we approximate $\mu_{i\gets j} \approx \mu_j$, $\chi_{i\gets j} \approx \chi_j$, and $\sum_{k \in \mathcal{N}_j \setminus i} \approx \sum_{k \in \mathcal{N}_j}$ in the BP equations, Eqs.~\eqref{eq:mu_ij_z1} and \eqref{eq:mu_i_z1}, and the SusP equations, Eqs.~\eqref{eq:BP_perc_chi_ij} and \eqref{eq:BP_perc_chi_i}. Under these approximations, the equations reduce to the naive mean-field approximation (MFA) equations:
\begin{align}
    \mu_{i} &\approx 1-\prod_{j\in\mathcal{N}_i}\left(1-p\,\mu_{j}\right), \label{eq:MFA_perc_m_i}
    \\
    \chi_i &\approx \left[ 1 + \sum_{j\in\mathcal{N}_i} \frac{p\,\chi_{j}}{1- p\,\mu_{j}} \right] \left(1-\mu_i\right). \label{eq:MFA_perc_chi_i}
\end{align}

In the source-node mean-field approximation (SNMFA), the source node $x$ always belongs to the infinite cluster, so we set $\mu_x = 1$ in Eq.~\eqref{eq:MFA_perc_m_i}: 
\begin{equation}
    \mu_{i}^{(x)} \approx 1 - (1-\delta_{ix}) \prod_{j\in\mathcal{N}_i}\left(1-p\,\mu_{j}^{(x)}\right).
\end{equation}

\section{Mean-field approximation for the Ising model}

For the MFA of the Ising model, we approximate $h_{i\gets j} \approx h_j$ and $\sum_{k \in \mathcal{N}_j \setminus i} \approx \sum_{k \in \mathcal{N}_j}$ in Eq.~\eqref{eq:h_relation}:
\begin{equation}\label{eq:MFA_h_relation}
        \tanh{\left(\beta h_{j}\right)} \approx \tanh{(\beta)} \cdot \tanh{\left(\sum_{k\in\mathcal{N}_j} \beta h_{k}\right)}.
\end{equation}
Since $m_i \approx \tanh{\left( \sum_{j\in\mathcal{N}_i} \beta h_j \right)}$, we can rewrite Eq.~\eqref{eq:MFA_h_relation} to obtain the naive MFA equation:
\begin{equation}\label{eq:MFA_mi_solution_from_h}
    m_i \approx \tanh{\left(\sum_{j\in\mathcal{N}_i} \tanh^{-1}\left( \tanh{(\beta)}\cdot m_j \right)\right)}.
\end{equation}
When $\beta \ll 1$ or $m_j \approx 1$ (which is generally valid since $m_j$ increases with $\beta$), this further approximates to the well-known form~\cite{Huang2021Statistical, Dorogovtsev2008Critical}:
\begin{equation}\label{eq:typical_local_MFA_si}
m_i \approx \tanh{\left(\sum_{j\in\mathcal{N}_i} \beta \, m_j \right)}.
\end{equation}
Numerical checks show that Eqs.~\eqref{eq:MFA_mi_solution_from_h} and \eqref{eq:typical_local_MFA_si} yield nearly identical order parameters on real networks, with differences negligible compared to the intrinsic error of the MFA. Therefore, we use Eq.~\eqref{eq:MFA_mi_solution_from_h} for all simulations.

For the MFA of magnetic susceptibility, we additionally approximate $q_{i\gets j}\approx q_{j}$ in the SusP equation (Eq.~\eqref{eq:susp_dh_relation}): 
\begin{align}\label{eq:mfa_dh_relation}
    q_{i} & \approx \frac{\tanh(\beta)}{\mathrm{sech}^2(\beta h_{i})} \left[ 1 - \frac{\tanh^2(\beta h_{i})}{\tanh^2(\beta)} \right] \Biggl( 1 + \sum_{j \in \mathcal{N}_i} q_{j} \Biggr) \\
    & \approx \tanh(\beta) \,\frac{1 - m_i^2 }{1-m_i^2 \tanh^2(\beta)} \Biggl( 1 + \sum_{j \in \mathcal{N}_i} q_{j} \Biggr)
\end{align}
where $m_i\approx {\tanh(\beta h_{i})}/{\tanh(\beta)}$ from Eqs.~\eqref{eq:MFA_h_relation}.
Applying the same approximation to Eq.~\eqref{eq:susp_local_chi}, the local susceptibility becomes
\begin{equation}\label{eq:mfa_local_chi}
    \chi_i \approx \beta \, (1 - m_i^2) \Biggl( 1 + \sum_{j \in \mathcal{N}_i} q_{j} \Biggr).
\end{equation}

In the SNMFA, we set $m_x=1$ for the source node $x$ in Eq.~\eqref{eq:MFA_mi_solution_from_h}:
\begin{equation}\label{eq:SNMFA_mi_solution_from_h}
    m_i^{(x)} \approx \delta_{ix} + (1-\delta_{ix})\tanh{\left(\sum_{j\in\mathcal{N}_i} \tanh^{-1}\left( \tanh{(\beta)}\cdot m_j^{(x)} \right)\right)}.
\end{equation}

\section{Monte Carlo Simulation Methods}

To benchmark the source-node belief propagation (SNBP) method, we perform Monte Carlo (MC) simulations for bond percolation and the ferromagnetic Ising model. We implement two approaches: conventional MC, preserving system symmetry for global observables, and source-node MC (SNMC), breaking symmetry by computing quantities relative to a source node $x$ (typically highest-degree). Observables are evaluated at 50 parameter values, with $p \in [0.01, 0.99]$ for percolation and $\beta = -({1}/{2}) \ln(1-p)$ (from $p = 1-e^{-2\beta}$ in the random cluster model~\cite{Wu1982Potts}) for the Ising model. The Ising model MC algorithm employs a hybrid approach, combining Wolff cluster updates with Metropolis single-spin flips, optimized through thermalization checks and adaptive measurement scaling. Below, we detail the algorithms in pseudocode, focusing on steps impacting the observables.

\begin{algorithm}[H]
\caption{Conventional MC for the Ising Model}
\begin{algorithmic}[1]
\State Initialize random spins $\sigma_i = \pm 1$ for all $i \in \mathcal{V}$.
\State Compute initial energy $E = -({1}/{2}) \sum_i \sigma_i \sum_{j \in \mathcal{N}_i} \sigma_j$; set smoothed energy $\tilde{E} = E$, $\alpha = 0.3$.
\State \textbf{Thermalization:}
\For{$i = 1$ to 100 macro-steps}
    \State Perform one Wolff update: select random vertex with neighbors; add like-aligned neighbors with probability $1 - \exp(-2\beta)$; flip cluster spins.
    \State Perform $50 \times N$ Metropolis single-spin flips with acceptance probability $\min(1, \exp(-2\beta \Delta E))$.
    \State Compute $m = \sum_i \sigma_i / N$, $|m|$, and update $E$, $\tilde{E} = \alpha E + (1-\alpha) \tilde{E}$ (skip smoothing if $i \leq 2$).
    \State \textbf{Stop early} if $|m| > 0.95$ for two consecutive macro-steps or $\tilde{E}_i - \tilde{E}_{i-1} > -0.001 |\tilde{E}_i|$ for six macro-steps ($i > 2$).
\EndFor
\State Compute $M_f$ as mean of last 7 (if $\geq 12$ macro-steps), 5 (if $\geq 10$ macro-steps), 3 (if $\geq 8$ macro-steps), or 2 $|m|$ values. 
\Comment{The number of iterations is determined via thermalization checks.}
\State \textbf{Measurement:}
\State Set base iterations $K = \text{round}(8 \times 10^5 / \sqrt{N})$; scale by $n=1$ if $M_f < 0.9$, else $n=5$. \Comment{The measurement frequency is adaptively scaled to enhance efficiency.}
\For{$k = 1$ to $\text{round}(K \times n)$}
    \State Perform one Wolff update and $50 \times N$ Metropolis flips.
    \State Compute $m = \sum_i \sigma_i /N$; store $|m|$ and $m^2$.
\EndFor
\State Compute $\langle |m| \rangle = \text{mean}(|m|)$, $\chi_{\rm true} = \beta N \langle m^2 \rangle$, $\chi_{\rm practical} = \beta N (\langle m^2 \rangle - \langle |m| \rangle^2)$.
\State Compute standard deviations (ddof=1) and standard errors for $\langle |m| \rangle$.
\State \Return $\langle |m| \rangle$, $\chi_{\rm true}$, $\chi_{\rm practical}$, with errors.
\end{algorithmic}
\end{algorithm}

\begin{algorithm}[H]
\caption{Source-Node MC for the Ising Model}
\begin{algorithmic}[1]
\State Follow Conventional MC (Algorithm 1), with additions:
\State During measurements, compute $m \sigma_x$ for source node $x$ and store alongside $|m|$ and $m^2$.
\State Compute $\langle m \sigma_x \rangle = \text{mean}(m \sigma_x)$, $\chi_{\rm source} = \beta N (\langle m^2 \rangle - \langle m \sigma_x \rangle^2)$, with standard deviation and error for $\langle m \sigma_x \rangle$.
\State \Return $\langle m \sigma_x \rangle$, $\chi_{\rm source}$, with errors.
\end{algorithmic}
\end{algorithm}

\begin{algorithm}[H]
\caption{Conventional MC for Bond Percolation}
\begin{algorithmic}[1]
\For{each $p \in [0.01, 0.99]$ (50 values)}
    \For{$r = 1$ to $4 \times 10^5$ realizations}
        \State Occupy edges $(i,j) \in \mathcal{E}$ with probability $p$.
        \State Identify connected components using BFS.
        \State Compute largest cluster size $|\mathcal{C}_{\max}|$, $S_1 = |\mathcal{C}_{\max}|/N$.
        \State Compute $\chi_{\rm true} = \sum_{\mathcal{C}} |\mathcal{C}|^2 /N$, $\chi_{\rm practical} = \sum_{\mathcal{C} \neq \mathcal{C}_{\max}} |\mathcal{C}|^2 /N$.
    \EndFor
    \State Compute means and standard deviations (ddof=1) for $S_1$, $\chi_{\rm true}$, $\chi_{\rm practical}$ over realizations.
\EndFor
\State \Return $\langle S_1 \rangle$, $\langle \chi_{\rm true} \rangle$, $\langle \chi_{\rm practical} \rangle$, with errors.
\end{algorithmic}
\end{algorithm}

\begin{algorithm}[H]
\caption{Source-Node MC for Bond Percolation}
\begin{algorithmic}[1]
\State Follow Conventional MC (Algorithm 3), with additions:
\For{each realization}
    \State Identify cluster $\mathcal{C}(x)$ containing source node $x$.
    \State Compute $S(x) = |\mathcal{C}(x)| / N$, $\chi_{\rm source} = \sum_{\mathcal{C} \neq \mathcal{C}(x)} |\mathcal{C}|^2 /N$.
\EndFor
\State Compute means and standard deviations (ddof=1) for $S(x)$, $\chi_{\rm source}$ over $4 \times 10^5$ realizations.
\State \Return $\langle S(x) \rangle$, $\langle \chi_{\rm source} \rangle$, with errors.
\end{algorithmic}
\end{algorithm}

\section{Definition of Errors}

This section provides the formal definitions for the error metrics $\Delta_{\mathrm{Method-MC}}$ used to benchmark the accuracy of various inference methods against MC simulations.
The error for a given method is calculated as the area between the method's OP curve and the MC OP curve, numerically integrated from 0 to 1. The specific definitions for each model are given below.

For bond percolation, the integration is over the bond occupation probability $p$:
\begin{equation}
    \Delta_{\mathrm{BP-MC}}  \equiv  \int_0^1 {\Big| \langle P_\infty \rangle_\mathrm{BP} - \langle S_1 \rangle_\mathrm{MC} \Big|} \, dp,
\end{equation}
\begin{equation}
    \Delta_{\mathrm{SNBP-MC}}  \equiv  \int_0^1 {\Big| \langle S(x) \rangle_\mathrm{SNBP} - \langle S_1 \rangle_\mathrm{MC} \Big|} \, dp,
\end{equation}
\begin{equation}
    \Delta_{\mathrm{MFA-MC}}  \equiv  \int_0^1 {\Big| \langle P_\infty \rangle_\mathrm{MFA} - \langle S_1 \rangle_\mathrm{MC} \Big|} \, dp.
\end{equation}

For the Ising model, the integration is over $p=1-e^{-2\beta}$:
\begin{equation}
    \Delta_{\mathrm{BP-MC}}  \equiv  \int_0^1 {\Big| \langle m \rangle_\mathrm{BP} - \langle |m| \rangle_\mathrm{MC} \Big|} \, dp,
\end{equation}
\begin{equation}
    \Delta_{\mathrm{SNBP-MC}}  \equiv  \int_0^1 {\Big| \langle m \rangle_\mathrm{SNBP} - \langle |m| \rangle_\mathrm{MC} \Big|} \, dp,
\end{equation}
\begin{equation}
    \Delta_{\mathrm{MFA-MC}}  \equiv  \int_0^1 {\Big| \langle m \rangle_\mathrm{MFA} - \langle |m| \rangle_\mathrm{MC} \Big|} \, dp.
\end{equation}

A smaller value of $\Delta_{\mathrm{Method-MC}}$ indicates a more accurate approximation. The comparative results for these errors across the 139 benchmark networks are presented in Fig.~4 and Supplemental Fig.~4.

\section{Network datasets}

Tables~S1 to S3 summarize the network datasets analyzed in Fig.~4 and Fig.~S4. All datasets were obtained from the Netzschleuder repository (\url{https://networks.skewed.de}). Each dataset can be directly loaded in Python using the graph-tool library with the command \texttt{g = gt.collection.ns["name"]}.
All original networks were preprocessed by extracting the largest connected component, removing self-loops, and removing parallel edges. The statistics shown correspond to these processed simple graphs. 

\setlength{\tabcolsep}{8pt}

\begin{table}
\begin{tabular}{lllllll}
\toprule
Network name & Domain & $N$ & $M$ & $c(\mathcal{G})$ & $\langle{k}\rangle$ & url \\
\midrule
adjnoun & Informational & 112 & 425 & 314 & 7.59 & \href{http://www-personal.umich.edu/~mejn/netdata/adjnoun.zip}{url} \\
baseball/user-provider & Social & 47 & 46 & 0 & 1.96 & \href{http://orgnet.com/steroids.html}{url} \\
blumenau\_drug & Biological & 75 & 181 & 107 & 4.83 & \href{https://github.com/rionbr/DDIBlumenau/tree/master/csv}{url} \\
board\_directors/net1m\_2002-05-01 & Social & 154 & 848 & 695 & 11.01 & \href{http://www.boardsandgender.com/data.php}{url} \\
board\_directors/net1m\_2005-06-01 & Social & 476 & 1836 & 1361 & 7.71 & \href{http://www.boardsandgender.com/data.php}{url} \\
board\_directors/net1m\_2008-07-01 & Social & 840 & 2700 & 1861 & 6.43 & \href{http://www.boardsandgender.com/data.php}{url} \\
board\_directors/net1m\_2011-08-01 & Social & 854 & 2745 & 1892 & 6.43 & \href{http://www.boardsandgender.com/data.php}{url} \\
board\_directors/net2m\_2005-05-01 & Social & 568 & 594 & 27 & 2.09 & \href{http://www.boardsandgender.com/data.php}{url} \\
board\_directors/net1m\_2002-06-01 & Social & 144 & 824 & 681 & 11.44 & \href{http://www.boardsandgender.com/data.php}{url} \\
celegans\_2019/hermaphrodite\_gap\_junction\_corrected & Biological & 460 & 1432 & 973 & 6.23 & \href{https://wormwiring.org/pages/adjacency.html}{url} \\
celegans\_2019/male\_gap\_junction\_corrected & Biological & 484 & 1597 & 1114 & 6.60 & \href{https://wormwiring.org/pages/adjacency.html}{url} \\
celegans\_2019/hermaphrodite\_gap\_junction & Biological & 460 & 1428 & 969 & 6.21 & \href{https://wormwiring.org/pages/adjacency.html}{url} \\
celegans\_2019/hermaphrodite\_gap\_junction\_synapse & Biological & 279 & 962 & 684 & 6.90 & \href{https://wormwiring.org/pages/adjacency.html}{url} \\
celegans\_2019/male\_gap\_junction & Biological & 484 & 1594 & 1111 & 6.59 & \href{https://wormwiring.org/pages/adjacency.html}{url} \\
celegans\_2019/male\_gap\_junction\_synapse & Biological & 298 & 1171 & 874 & 7.86 & \href{https://wormwiring.org/pages/adjacency.html}{url} \\
celegans\_interactomes/BPmaps & Biological & 345 & 400 & 56 & 2.32 & \href{http://interactome.dfci.harvard.edu/C_elegans/index.php?page=download}{url} \\
celegans\_interactomes/Genetic & Biological & 683 & 1543 & 861 & 4.52 & \href{http://interactome.dfci.harvard.edu/C_elegans/index.php?page=download}{url} \\
celegans\_interactomes/LCI & Biological & 117 & 123 & 7 & 2.10 & \href{http://interactome.dfci.harvard.edu/C_elegans/index.php?page=download}{url} \\
ceo\_club & Social & 40 & 95 & 56 & 4.75 & \href{http://konect.cc/networks/brunson_club-membership}{url} \\
contiguous\_usa & Transportation & 49 & 107 & 59 & 4.37 & \href{http://konect.cc/networks/contiguous-usa}{url} \\
copenhagen/fb\_friends & Social & 800 & 6418 & 5619 & 16.05 & \href{https://doi.org/10.6084/m9.figshare.7267433}{url} \\
dolphins & Social & 62 & 159 & 98 & 5.13 & \href{http://www-personal.umich.edu/~mejn/netdata/}{url} \\
edit\_wikibooks/la & Informational & 740 & 1051 & 312 & 2.84 & \href{http://konect.cc/networks/edit-enwikibooks}{url} \\
edit\_wikibooks/za & Informational & 46 & 45 & 0 & 1.96 & \href{http://konect.cc/networks/edit-enwikibooks}{url} \\
edit\_wikibooks/af & Informational & 625 & 724 & 100 & 2.32 & \href{http://konect.cc/networks/edit-enwikibooks}{url} \\
edit\_wikibooks/cv & Informational & 585 & 646 & 62 & 2.21 & \href{http://konect.cc/networks/edit-enwikibooks}{url} \\
edit\_wikibooks/lb & Informational & 84 & 84 & 1 & 2.00 & \href{http://konect.cc/networks/edit-enwikibooks}{url} \\
edit\_wikibooks/oc & Informational & 668 & 845 & 178 & 2.53 & \href{http://konect.cc/networks/edit-enwikibooks}{url} \\
edit\_wikiquote/co & Informational & 126 & 126 & 1 & 2.00 & \href{http://konect.cc/networks/edit-enwikiquote}{url} \\
edit\_wikiquote/mr & Informational & 799 & 963 & 165 & 2.41 & \href{http://konect.cc/networks/edit-enwikiquote}{url} \\
edit\_wikiquote/ga & Informational & 74 & 100 & 27 & 2.70 & \href{http://konect.cc/networks/edit-enwikiquote}{url} \\
edit\_wikiquote/kk & Informational & 50 & 49 & 0 & 1.96 & \href{http://konect.cc/networks/edit-enwikiquote}{url} \\
edit\_wikiquote/vo & Informational & 42 & 62 & 21 & 2.95 & \href{http://konect.cc/networks/edit-enwikiquote}{url} \\
edit\_wikiquote/am & Informational & 246 & 251 & 6 & 2.04 & \href{http://konect.cc/networks/edit-enwikiquote}{url} \\
edit\_wiktionary/aa & Informational & 32 & 51 & 20 & 3.19 & \href{http://konect.cc/networks/edit-frwiktionary}{url} \\
edit\_wiktionary/ab & Informational & 170 & 176 & 7 & 2.07 & \href{http://konect.cc/networks/edit-frwiktionary}{url} \\
edit\_wiktionary/rm & Informational & 75 & 74 & 0 & 1.97 & \href{http://konect.cc/networks/edit-frwiktionary}{url} \\
edit\_wiktionary/rn & Informational & 42 & 41 & 0 & 1.95 & \href{http://konect.cc/networks/edit-frwiktionary}{url} \\
edit\_wiktionary/dv & Informational & 967 & 1972 & 1006 & 4.08 & \href{http://konect.cc/networks/edit-frwiktionary}{url} \\
edit\_wiktionary/mh & Informational & 34 & 35 & 2 & 2.06 & \href{http://konect.cc/networks/edit-frwiktionary}{url} \\
ego\_social/facebook\_0 & Social & 324 & 2514 & 2191 & 15.52 & \href{http://snap.stanford.edu/data/egonets-Facebook.html}{url} \\
ego\_social/gplus\_117866881767579360121 & Social & 117 & 720 & 604 & 12.31 & \href{http://snap.stanford.edu/data/egonets-Facebook.html}{url} \\
ego\_social/gplus\_114336431216099933033 & Social & 455 & 4540 & 4086 & 19.96 & \href{http://snap.stanford.edu/data/egonets-Facebook.html}{url} \\
ego\_social/facebook\_3437 & Social & 532 & 4812 & 4281 & 18.09 & \href{http://snap.stanford.edu/data/egonets-Facebook.html}{url} \\
ego\_social/facebook\_3980 & Social & 44 & 138 & 95 & 6.27 & \href{http://snap.stanford.edu/data/egonets-Facebook.html}{url} \\
ego\_social/facebook\_686 & Social & 168 & 1656 & 1489 & 19.71 & \href{http://snap.stanford.edu/data/egonets-Facebook.html}{url} \\
elite & Social & 44 & 99 & 56 & 4.50 & \href{http://konect.cc/networks/brunson_corporate-leadership}{url} \\
eu\_airlines & Transportation & 417 & 2953 & 2537 & 14.16 & \href{https://manliodedomenico.com/data.php}{url} \\
eu\_procurements\_alt/IE\_2013 & Economic & 997 & 1040 & 44 & 2.09 & \href{https://zenodo.org/record/3627216#.XivHNy2ZNGV}{url} \\
eu\_procurements\_alt/SK\_2008 & Economic & 660 & 773 & 114 & 2.34 & \href{https://zenodo.org/record/3627216#.XivHNy2ZNGV}{url} \\
\bottomrule
\end{tabular}
\caption{Network datasets used in Fig.~4 and S4 (Part 1 of 3). For each network, we list its name, domain, number of nodes $N$, number of edges $M$, cyclomatic number $c(\mathcal{G})\equiv M-N+1$, mean degree $\langle k \rangle$, and data source (URLs are provided as hyperlinks).}
\label{tab:network_datasets_part1}
\end{table}

\begin{table}
\begin{tabular}{lllllll}
\toprule
Network name & Domain & $N$ & $M$ & $c(\mathcal{G})$ & $\langle{k}\rangle$ & url \\
\midrule
eu\_procurements\_alt/SK\_2009 & Economic & 696 & 795 & 100 & 2.28 & \href{https://zenodo.org/record/3627216#.XivHNy2ZNGV}{url} \\
eu\_procurements\_alt/SK\_2010 & Economic & 593 & 716 & 124 & 2.41 & \href{https://zenodo.org/record/3627216#.XivHNy2ZNGV}{url} \\
eu\_procurements\_alt/EE\_2008 & Economic & 480 & 540 & 61 & 2.25 & \href{https://zenodo.org/record/3627216#.XivHNy2ZNGV}{url} \\
eu\_procurements\_alt/PT\_2008 & Economic & 692 & 776 & 85 & 2.24 & \href{https://zenodo.org/record/3627216#.XivHNy2ZNGV}{url} \\
facebook\_friends & Social & 329 & 1954 & 1626 & 11.88 & \href{https://github.com/benmaier/BFMaierFBnetwork}{url} \\
facebook\_organizations/S1 & Social & 320 & 2369 & 2050 & 14.81 & \href{https://data4goodlab.github.io/MichaelFire/#section3}{url} \\
facebook\_organizations/S2 & Social & 165 & 726 & 562 & 8.80 & \href{https://data4goodlab.github.io/MichaelFire/#section3}{url} \\
football & Social & 115 & 613 & 499 & 10.66 & \href{http://www-personal.umich.edu/~mejn/netdata}{url} \\
football\_tsevans & Social & 115 & 613 & 499 & 10.66 & \href{https://figshare.com/articles/American_College_Football_Network_Files/93179}{url} \\
fullerene\_structures/C260 & Biological & 260 & 390 & 131 & 3.00 & \href{https://doi.org/10.6084/m9.figshare.5700832}{url} \\
fullerene\_structures/C320 & Biological & 320 & 480 & 161 & 3.00 & \href{https://doi.org/10.6084/m9.figshare.5700832}{url} \\
fullerene\_structures/C540 & Biological & 540 & 810 & 271 & 3.00 & \href{https://doi.org/10.6084/m9.figshare.5700832}{url} \\
fullerene\_structures/C960 & Biological & 960 & 1440 & 481 & 3.00 & \href{https://doi.org/10.6084/m9.figshare.5700832}{url} \\
fullerene\_structures/C180 & Biological & 180 & 270 & 91 & 3.00 & \href{https://doi.org/10.6084/m9.figshare.5700832}{url} \\
fullerene\_structures/C240 & Biological & 240 & 360 & 121 & 3.00 & \href{https://doi.org/10.6084/m9.figshare.5700832}{url} \\
game\_thrones & Social & 107 & 352 & 246 & 6.58 & \href{http://www.macalester.edu/~abeverid/data/stormofswords.csv}{url} \\
human\_brains/BNU1\_0025864\_1\_DTI\_CPAC200 & Biological & 200 & 1832 & 1633 & 18.32 & \href{https://awesome.cs.jhu.edu/graph-services/download/}{url} \\
human\_brains/Jung2015\_M87125989\_1\_DTI\_CPAC200 & Biological & 200 & 1760 & 1561 & 17.60 & \href{https://awesome.cs.jhu.edu/graph-services/download/}{url} \\
human\_brains/BNU1\_0025864\_1\_DTI\_DS00071 & Biological & 67 & 495 & 429 & 14.78 & \href{https://awesome.cs.jhu.edu/graph-services/download/}{url} \\
human\_brains/Jung2015\_M87125989\_1\_DTI\_DS00071 & Biological & 68 & 466 & 399 & 13.71 & \href{https://awesome.cs.jhu.edu/graph-services/download/}{url} \\
human\_brains/MRN1313\_FDL\_1\_DTI\_slab907 & Biological & 848 & 1887 & 1040 & 4.45 & \href{https://awesome.cs.jhu.edu/graph-services/download/}{url} \\
human\_brains/MRN1313\_S9X\_1\_DTI\_CPAC200 & Biological & 200 & 1563 & 1364 & 15.63 & \href{https://awesome.cs.jhu.edu/graph-services/download/}{url} \\
interactome\_pdz & Biological & 161 & 209 & 49 & 2.60 & \href{http://konect.cc/networks/maayan-pdzbase}{url} \\
karate/77 & Social & 34 & 77 & 44 & 4.53 & \href{https://aaronclauset.github.io/datacode.htm}{url} \\
karate/78 & Social & 34 & 78 & 45 & 4.59 & \href{https://aaronclauset.github.io/datacode.htm}{url} \\
kegg\_metabolic/aae & Biological & 880 & 2296 & 1417 & 5.22 & \href{http://santafe.edu/~aaronc/data/kegg2006_metabolic.zip}{url} \\
kegg\_metabolic/fnu & Biological & 993 & 2455 & 1463 & 4.94 & \href{http://santafe.edu/~aaronc/data/kegg2006_metabolic.zip}{url} \\
kegg\_metabolic/mpe & Biological & 546 & 1289 & 744 & 4.72 & \href{http://santafe.edu/~aaronc/data/kegg2006_metabolic.zip}{url} \\
kegg\_metabolic/sso & Biological & 992 & 2455 & 1464 & 4.95 & \href{http://santafe.edu/~aaronc/data/kegg2006_metabolic.zip}{url} \\
kegg\_metabolic/afu & Biological & 861 & 2011 & 1151 & 4.67 & \href{http://santafe.edu/~aaronc/data/kegg2006_metabolic.zip}{url} \\
kegg\_metabolic/hal & Biological & 783 & 1986 & 1204 & 5.07 & \href{http://santafe.edu/~aaronc/data/kegg2006_metabolic.zip}{url} \\
lesmis & Social & 77 & 254 & 178 & 6.60 & \href{http://www-personal.umich.edu/~mejn/netdata}{url} \\
london\_transport & Transportation & 369 & 430 & 62 & 2.33 & \href{https://manliodedomenico.com/data.php}{url} \\
malaria\_genes/HVR\_1 & Biological & 307 & 2812 & 2506 & 18.32 & \href{https://github.com/dblarremore/data_malaria_PLOSCompBiology_2013}{url} \\
malaria\_genes/HVR\_2 & Biological & 112 & 743 & 632 & 13.27 & \href{https://github.com/dblarremore/data_malaria_PLOSCompBiology_2013}{url} \\
malaria\_genes/HVR\_4 & Biological & 183 & 933 & 751 & 10.20 & \href{https://github.com/dblarremore/data_malaria_PLOSCompBiology_2013}{url} \\
malaria\_genes/HVR\_5 & Biological & 298 & 2684 & 2387 & 18.01 & \href{https://github.com/dblarremore/data_malaria_PLOSCompBiology_2013}{url} \\
marvel\_partnerships & Social & 181 & 224 & 44 & 2.48 & \href{http://ix.io/1omF}{url} \\
mist/genetic\_mouse & Biological & 214 & 252 & 39 & 2.36 & \href{https://fgrtools.hms.harvard.edu/MIST/index.jsp}{url} \\
mist/ppi\_zebrafish & Biological & 197 & 225 & 29 & 2.28 & \href{https://fgrtools.hms.harvard.edu/MIST/index.jsp}{url} \\
moviegalaxies/364 & Social & 39 & 105 & 67 & 5.38 & \href{https://dataverse.harvard.edu/dataset.xhtml?persistentId=doi:10.7910/DVN/T4HBA3}{url} \\
moviegalaxies/642 & Social & 30 & 99 & 70 & 6.60 & \href{https://dataverse.harvard.edu/dataset.xhtml?persistentId=doi:10.7910/DVN/T4HBA3}{url} \\
moviegalaxies/777 & Social & 37 & 82 & 46 & 4.43 & \href{https://dataverse.harvard.edu/dataset.xhtml?persistentId=doi:10.7910/DVN/T4HBA3}{url} \\
moviegalaxies/92 & Social & 71 & 154 & 84 & 4.34 & \href{https://dataverse.harvard.edu/dataset.xhtml?persistentId=doi:10.7910/DVN/T4HBA3}{url} \\
moviegalaxies/235 & Social & 39 & 137 & 99 & 7.03 & \href{https://dataverse.harvard.edu/dataset.xhtml?persistentId=doi:10.7910/DVN/T4HBA3}{url} \\
moviegalaxies/643 & Social & 39 & 197 & 159 & 10.10 & \href{https://dataverse.harvard.edu/dataset.xhtml?persistentId=doi:10.7910/DVN/T4HBA3}{url} \\
netscience & Social & 379 & 914 & 536 & 4.82 & \href{http://www-personal.umich.edu/~mejn/netdata/}{url} \\
plant\_pol\_kato & Biological & 768 & 1205 & 438 & 3.14 & \href{https://iwdb.nceas.ucsb.edu/html/kato_1990.html}{url} \\
plant\_pol\_vazquez/All sites pooled & Biological & 104 & 164 & 61 & 3.15 & \href{https://iwdb.nceas.ucsb.edu/html/vazquez_2002.html}{url} \\
plant\_pol\_vazquez/Arroyo Goye & Biological & 35 & 41 & 7 & 2.34 & \href{https://iwdb.nceas.ucsb.edu/html/vazquez_2002.html}{url} \\
\bottomrule
\end{tabular}
\caption{Network datasets used in Fig.~4 and S4 (Part 2 of 3).}
\label{tab:network_datasets_part2}
\end{table}

\begin{table}
\begin{tabular}{lllllll}
\toprule
Network name & Domain & $N$ & $M$ & $c(\mathcal{G})$ & $\langle{k}\rangle$ & url \\
\midrule
plant\_pol\_vazquez/Cerro Lopez & Biological & 40 & 44 & 5 & 2.20 & \href{https://iwdb.nceas.ucsb.edu/html/vazquez_2002.html}{url} \\
plant\_pol\_vazquez/Llao Llao & Biological & 32 & 36 & 5 & 2.25 & \href{https://iwdb.nceas.ucsb.edu/html/vazquez_2002.html}{url} \\
plant\_pol\_vazquez/Mascardi (c) & Biological & 30 & 34 & 5 & 2.27 & \href{https://iwdb.nceas.ucsb.edu/html/vazquez_2002.html}{url} \\
plant\_pol\_vazquez/Mascardi (nc) & Biological & 39 & 48 & 10 & 2.46 & \href{https://iwdb.nceas.ucsb.edu/html/vazquez_2002.html}{url} \\
polbooks & Informational & 105 & 441 & 337 & 8.40 & \href{http://www-personal.umich.edu/~mejn/netdata}{url} \\
product\_space/HS & Economic & 866 & 2532 & 1667 & 5.85 & \href{http://www.michelecoscia.com/?page_id=223}{url} \\
product\_space/SITC & Economic & 774 & 1779 & 1006 & 4.60 & \href{http://www.michelecoscia.com/?page_id=223}{url} \\
revolution & Social & 141 & 160 & 20 & 2.27 & \href{http://konect.cc/networks/brunson_revolution}{url} \\
route\_views/19990829 & Technological & 103 & 239 & 137 & 4.64 & \href{http://snap.stanford.edu/data/as.html}{url} \\
route\_views/19981229 & Technological & 493 & 1145 & 653 & 4.65 & \href{http://snap.stanford.edu/data/as.html}{url} \\
route\_views/19981230 & Technological & 522 & 1198 & 677 & 4.59 & \href{http://snap.stanford.edu/data/as.html}{url} \\
route\_views/19981231 & Technological & 512 & 1181 & 670 & 4.61 & \href{http://snap.stanford.edu/data/as.html}{url} \\
route\_views/19990101 & Technological & 531 & 1217 & 687 & 4.58 & \href{http://snap.stanford.edu/data/as.html}{url} \\
route\_views/19990102 & Technological & 541 & 1233 & 693 & 4.56 & \href{http://snap.stanford.edu/data/as.html}{url} \\
sp\_high\_school/facebook & Social & 156 & 1437 & 1282 & 18.42 & \href{http://www.sociopatterns.org/datasets/high-school-contact-and-friendship-networks/}{url} \\
student\_cooperation & Social & 141 & 256 & 116 & 3.63 & \href{https://data4goodlab.github.io/MichaelFire/#section3}{url} \\
terrorists\_911 & Social & 62 & 152 & 91 & 4.90 & \href{https://aaronclauset.github.io/datacode.htm}{url} \\
train\_terrorists & Social & 64 & 243 & 180 & 7.59 & \href{http://konect.cc/networks/moreno_train}{url} \\
tree-of-life/360911 & Biological & 498 & 1949 & 1452 & 7.83 & \href{http://snap.stanford.edu/tree-of-life/data.html}{url} \\
tree-of-life/469613 & Biological & 409 & 856 & 448 & 4.19 & \href{http://snap.stanford.edu/tree-of-life/data.html}{url} \\
tree-of-life/5762 & Biological & 591 & 3356 & 2766 & 11.36 & \href{http://snap.stanford.edu/tree-of-life/data.html}{url} \\
tree-of-life/715226 & Biological & 67 & 113 & 47 & 3.37 & \href{http://snap.stanford.edu/tree-of-life/data.html}{url} \\
tree-of-life/9986 & Biological & 34 & 450 & 417 & 26.47 & \href{http://snap.stanford.edu/tree-of-life/data.html}{url} \\
tree-of-life/1000570 & Biological & 167 & 248 & 82 & 2.97 & \href{http://snap.stanford.edu/tree-of-life/data.html}{url} \\
ugandan\_village/friendship-1 & Social & 202 & 547 & 346 & 5.42 & \href{https://www.repository.cam.ac.uk/handle/1810/270256?show=full}{url} \\
ugandan\_village/friendship-14 & Social & 124 & 525 & 402 & 8.47 & \href{https://www.repository.cam.ac.uk/handle/1810/270256?show=full}{url} \\
ugandan\_village/friendship-3 & Social & 192 & 1060 & 869 & 11.04 & \href{https://www.repository.cam.ac.uk/handle/1810/270256?show=full}{url} \\
ugandan\_village/friendship-8 & Social & 369 & 1753 & 1385 & 9.50 & \href{https://www.repository.cam.ac.uk/handle/1810/270256?show=full}{url} \\
ugandan\_village/health-advice\_12 & Social & 218 & 534 & 317 & 4.90 & \href{https://www.repository.cam.ac.uk/handle/1810/270256?show=full}{url} \\
ugandan\_village/health-advice\_17 & Social & 63 & 146 & 84 & 4.63 & \href{https://www.repository.cam.ac.uk/handle/1810/270256?show=full}{url} \\
unicodelang & Informational & 858 & 1249 & 392 & 2.91 & \href{http://konect.cc/networks/unicodelang}{url} \\
urban\_streets/brasilia & Transportation & 179 & 230 & 52 & 2.57 & \href{https://www.complex-networks.net/datasets.html#chap8}{url} \\
urban\_streets/irvine2 & Transportation & 178 & 189 & 12 & 2.12 & \href{https://www.complex-networks.net/datasets.html#chap8}{url} \\
urban\_streets/new-delhi & Transportation & 252 & 328 & 77 & 2.60 & \href{https://www.complex-networks.net/datasets.html#chap8}{url} \\
urban\_streets/richmond & Transportation & 697 & 1084 & 388 & 3.11 & \href{https://www.complex-networks.net/datasets.html#chap8}{url} \\
urban\_streets/seoul & Transportation & 869 & 1307 & 439 & 3.01 & \href{https://www.complex-networks.net/datasets.html#chap8}{url} \\
urban\_streets/walnut-creek & Transportation & 169 & 196 & 28 & 2.32 & \href{https://www.complex-networks.net/datasets.html#chap8}{url} \\
wiki\_science & Informational & 677 & 6517 & 5841 & 19.25 & \href{https://figshare.com/articles/A_Wikipedia_Based_Map_of_Science/11638932}{url} \\
windsurfers & Social & 43 & 336 & 294 & 15.63 & \href{http://konect.cc/networks/moreno_beach}{url} \\
\bottomrule
\end{tabular}
\caption{Network datasets used in Fig.~4 and S4 (Part 3 of 3).}
\label{tab:network_datasets_part3}
\end{table}



\end{document}